\documentclass[prb,twocolumn,letterpaper,showpacs,superscriptaddress]{revtex4}
\usepackage{graphicx,latexsym,amsfonts,amsmath,amssymb,bm}
\usepackage{dcolumn} 
\begin{document}
\title{Asymptotic Capture-Number and Island-Size Distributions \\
for One-Dimensional Irreversible Submonolayer Growth}
\date{\today}
\author{J. G. Amar}
\email{jamar@physics.utoledo.edu}
\affiliation{Department of Physics \& Astronomy,
University of Toledo, Toledo, OH 43606, USA}
\author{M. N. Popescu}
\email{popescu@mf.mpg.de}
\affiliation{
Max-Planck-Institut f\"ur Metallforschung, Heisenbergstr. 3,
D-70569 Stuttgart, Germany}
\affiliation{Institut f\"ur Theoretische und Angewandte Physik,
Universit\"at Stuttgart, Pfaffenwaldring 57, D-70569 Stuttgart,
Germany}

\begin{abstract}
  Using a  set of evolution equations  [J.G. Amar {\it  et al}, Phys. 
Rev. Lett. {\bf 86}, 3092 (2001)]  for the   average gap-size between 
islands, we calculate analytically the  asymptotic scaled 
capture-number distribution (CND) for one-dimensional irreversible 
submonolayer growth of point islands.   The predicted asymptotic CND 
is in reasonably good agreement with kinetic Monte-Carlo (KMC) 
results and leads to a \textit{non-divergent asymptotic} scaled 
island-size distribution (ISD). We then show  that  a  slight 
modification of   our analytical form leads to an analytic expression 
for the  asymptotic CND and a resulting asymptotic ISD which are in 
excellent agreement with KMC simulations. We also  show that in the 
asymptotic limit the self-averaging property of the capture zones 
holds exactly while the asymptotic scaled gap distribution is equal 
to the scaled CND. \end{abstract}


\maketitle

Recently, considerable theoretical effort has been carried out
towards a better understanding of the scaling properties of the
island-size distribution
  in
submonolayer  epitaxial growth.
\cite{scaling,BErapid,re_sc_1d, JMF_recent, APFsurf, Vvedensky,
recent, BE_recent}
For example, in the pre-coalescence regime the island size
distribution $N_s(\theta)$ (where $N_s$ is the number of islands of
size $s$ at coverage $\theta$) satisfies the scaling form,
\cite{scaling}\begin{equation}
N_s(\theta)= \frac{\theta}{S^2}f\left(\frac{s}{S}\right),
\label{scaling_f}
\end{equation}
where $S$ is the average island size, and the scaling function $f(u)$ depends
on  the critical island size and on the island morphology.\cite{scaling}

One of the standard tools used in these studies is the
rate-equation (RE) approach which involves a set of deterministic
coupled reaction-diffusion equations describing the coverage-dependence
of
$N_s(\theta)$  through a set of rate-coefficients usually called
capture numbers. \cite{Smo,Venables}
For the  irreversible growth of point islands,
rate-equations valid in the pre-coalescence regime may be written
in the form
\begin{subequations}
\begin {equation}
\label{monomer}
\frac{d N_1}{d \theta} = 1 - 2 R \sigma_1 N_1^2 -
R N_1 \sum_{s \geq 2} \sigma_s N_s
\end{equation}
\begin{equation}
\label{cluster}
~~~~~~~~\frac{d N_s}{d \theta} =  R N_1 \left( \sigma_{s-1} N_{s-1}  -
\sigma_s N_s \right), ~~\textrm{for}~ s \geq 2
\end{equation}
\label{N,N1}
\end{subequations}
where the capture numbers  ${\sigma}_s$ ($\sigma_1$) correspond to
the {\it average} capture rate of diffusing monomers by islands of
size $s$ (monomers) and $R = D/ F$ is the ratio of the monomer
diffusion rate to the deposition rate.
Accordingly, the  central problem in using the RE approach is the
determination of the average capture numbers $\sigma_s(\theta)$ and
the corresponding  capture number distribution (CND).

Recently, we
have developed
a self-consistent rate-equation approach to irreversible submonolayer
growth in which  correlations between the size of an island and the
corresponding {\it average}  capture zone are explicitly taken into
account in order to accurately predict the scaled island size and
capture number distributions.\cite{JMF_recent,APFsurf}   Our method
involves numerical integration of the island-density RE's (2) along
with the analytical solution of a set of approximate evolution
equations for the average  Voronoi-area surrounding an island of
size $s$, and is based on the following two assumptions:  (1) the
average capture area per freshly nucleated
dimer is
proportional (before rescaling) to the average
area per island and (2) the combined
effects of the preferred nucleation of dimers in large capture zones and the
preferred ``break-up'' of large capture zones due to nucleation may be
approximated by a uniform rescaling of the average capture zone of each island.
\cite{JMF_recent,APFsurf} Using this approach, we have obtained
numerical results for the scaled  capture number and island-size
distributions which agree well with kinetic Monte Carlo (KMC)
simulations in both one and two dimensions  over a wide range of
experimentally relevant values of $D/F$ ($D/F = 10^5 - 10^9$).

Recently, it has been argued \cite{Coment03} that because our method
does not explicitly take into account spatial fluctuations in the
nucleation, then it must lead to a diverging ISD in the asymptotic
limit corresponding to infinite $D/F$.  However, as shown by Bartelt
and Evans,\cite{BErapid} in the asymptotic limit the scaled ISD is
related to the scaled CND as,
\begin{equation}
f(u) = f(0) \exp \left[\,
\int_0^u dx \,\frac{2z-1-C'(x)}{C(x)-z ~x} \,
\right],
\label{asymptotic_f}
\end{equation}
where $C(s/S) = \sigma_s/\sigma_{av}$ is the scaled CND, $z$ is the
dynamical exponent describing the dependence of the average island
size on coverage ($S \sim \theta^z$), and $f(0)$ is determined by the
normalization condition,
\begin{equation}
\label{sumrule}
\int_0^\infty du~ f(u) = 1.
\end{equation}
As pointed out in
Ref. \onlinecite{BErapid},  Eq. \ref{asymptotic_f}
implies that
if   $C(u) > z u$ then no divergence will occur.  However, if $C(u)$ crosses
$z u$  at some value $u_c$  then the ISD will be cut off at $u_c$ 
(i.e. $f(u) = 0$ for $u \ge u_c$) if 
$C'(u_c) > 2 z - 1$ while a 
divergence in
the ISD    will occur  if $C'(u_c) < 2 z - 1$.  An example
of  a divergent asymptotic ISD is the 
usual mean-field theory with 
$C(u) = 1$.    
Thus the question of whether or not our
method  leads to a divergence in the asymptotic limit is entirely
determined by the   asymptotic scaled CND.

In this Rapid Communication we rigorously address the question of the
asymptotic behavior obtained using our method by analytically deriving the
asymptotic scaled CND along with  the corresponding asymptotic
scaled ISD for the case of irreversible growth of point islands in
one-dimension.
We find that, contrary to the claims in Ref.
\onlinecite{Coment03},
our method leads to a {\it non-divergent,
asymptotic} ISD as well as to an asymptotic scaled CND which is close
to that obtained in simulations.    We then show  that by slightly
modifying  our original analytical form for $C(u)$,
an  improved  analytical  expression for    the asymptotic scaled CND may
be obtained which is in excellent agreement with KMC simulations.
The resulting analytic expression   leads to a scaled ISD which is
also in excellent agreement with KMC simulations.
We also 
demonstrate that the asymptotic scaled gap distribution is identical 
to the scaled CND and as a result the self-averaging property of the 
capture zones holds exactly.

For clarity, we briefly review our method and its application to the case of
irreversible growth of point islands in one-dimension.\cite{APFsurf}
In this case, we have shown that the {\it local} capture-number
$\tilde \sigma(y)$ for an island with gap-size $y$ (corresponding to the
distance to the nearest island)
is
\begin{equation}
\label{sigmalocal}
\tilde \sigma(y;\theta) = \frac{2\xi_1}{\xi^2}~ \tanh \left(
\frac{y}{2\xi_1}\right),
\end{equation}
where the monomer capture length $\xi$ and the nucleation length $\xi_1$
are defined as
\begin{equation}
\label{xi1_xi}
2 \sigma_1 N_1 = 1/\xi_1^2,~~~~~1/\xi_1^2 + \sum_{s \ge 2} 
~\sigma_s~N_s = 1/\xi^2.
\end{equation}
  From the evolution equations for the distributions of gap lengths,
the average ``gap-length" $\tilde y_s$ (before rescaling) corresponding
to an island of size $s$ may be obtained as the solution of the equation,
\begin{equation}
\label{xytransform}
s - 2 =
\int_{\theta_y}^\theta R \, N_1(\phi)\, \tilde \sigma(\tilde y_s;\phi)\,d\phi.
\end{equation}
The coverage $\theta_y$ is defined by
$\tilde y_s = b ~Y(\theta_y)$
where  $b$ is the
proportionality factor (before rescaling)  between the average
gap-length of a freshly nucleated dimer and the average gap-length of
all islands,  $Y(\theta) = 1/N(\theta)$ is the
average gap-length at coverage $\theta$, and $N(\theta) = \sum_{s \ge 2}
N_s(\theta)$ is the average
island density.
Physically, the integral in Eq.~\ref{xytransform} may be interpreted
as corresponding to the  {\it average} number of particles added
to the dimer since it was formed at coverage $\theta_y$, neglecting any
change in
the capture zone due to nucleation and break-up.
To include the effects of break-up,
the gap lengths are rescaled
by a rescaling factor
$a = 1/\sum_{s \ge 2} N_s~ \tilde y_s$
so that
  the average gap length for an island of size $s$ is given by $y_s = 
a~\tilde y_s$
while the  capture number is given by  $\sigma_s = \tilde \sigma(y_s)$.

We now consider the
asymptotic limit
corresponding to infinite $D/F$.
In this limit, and assuming that
$\theta >> \theta_x$ (where $\theta_x \sim R^{-1/3}$ corresponds to the
coverage at which the island-density equals  the monomer density),
Eq.~\ref{sigmalocal} may be rewritten as,
\begin{equation}\label{localav}
\tilde \sigma(y;\theta) \simeq  y/\xi^2.
\end{equation}
This  implies that in the asymptotic limit the scaled gap-distribution
$B(s/S) = y_s/Y$ is \textit{identical} to the corresponding scaled capture
number distribution $C(s/S)$. This result also demonstrates the
``self-averaging property" of the capture zones,
i.e.
in the asymptotic limit  the average capture number $\sigma_s$
is  exactly  equal  to the {\it local } capture
number evaluated at the average capture zone for an island of size $s$.

In the
asymptotic limit that $\theta >> \theta_x$ one also has $dN_1/d\theta = 1
- R N_1/\xi^2 \simeq 0$,
which implies that $R N_1/\xi^2 = 1$. Using these results in
Eq.~\ref{xytransform} and assuming that also $\theta_y >>
\theta_x$,\cite{largey}
we obtain
\begin{equation}
\label{s-2}
s - 2 = \tilde y_s ~(\theta - \theta_y).
\end{equation}
Since in the asymptotic limit $S \gg 1$, this may be rewritten as
\begin{equation}
\label{xyfix}
u = s/S = (\tilde y_s \theta/S) (1 -  \theta_y/\theta).
\end{equation}

In the asymptotic limit, one has $\theta/S = N = 1/Y$.
Using $\tilde 
y_s = b/N(\theta_y)$ and $N(\theta_y) \sim \theta_y^{1/4}$,
$N(\theta) \sim \theta^{1/4}$ (since $\theta_y,~\theta \gg \theta_x$ and
$z = 3/4$ for irreversible growth of point-islands in one-dimension),
one has $(\theta_y/\theta) = (b Y/\tilde y_s)^4$. Thus, the
  asymptotic gap-distribution
before rescaling
will satisfy,
\begin{equation}
\label{first}
u = \tilde B(u) - b^4/\tilde B(u)^3,
\end{equation}
where $\tilde B(u) = \tilde y_s/Y$.
Rescaling,  we obtain for the asymptotic scaled gap
distribution $B(u) = a~\tilde B(u)$
  the equation
\begin{equation}
\label{firstfix}
u = B(u)/a - b^4 ~[B(u)/a]^{-3},
\end{equation}
where the rescaling factor $a$ and the proportionality factor $ b$
must satisfy the
requirement that
\begin{equation}\label{crossintegral}\int_0^\infty B(u) f(u) ~du =
1\end{equation}
It is easy to see that for any positive constants $a$ and $b$, Eq. 12
has a unique positive solution for every positive $u$ while
for $a \ge z$ one has $B(u) > z u$.  Similarly, one may show that
for $a < 2/3$ the ISD predicted by Eq. 3 will diverge
while   for $2/3 < a < z $ there will be a cut-off at $u_c$, i.e. 
$f(u > u_c) = 0$.

We now consider  the  case   $b = 1$ as was assumed in our
numerical calculations for finite $D/F$.\cite{JMF_recent,APFsurf}  Eq.
\ref{asymptotic_f} along with the requirement that $f(u) \rightarrow
0$ rapidly for large $u$, indicates that the asymptotic $B(u)$ should
satisfy $B(u) \simeq z u$ for large $u$. \cite{BE_recent}  Applied to
Eq. \ref{firstfix} this implies that  $a = z$ and thus $C(0) = B(0) =
3/4$, in very good agreement with our KMC simulation results.  This
leads to the following expression,
\begin{equation}\label{eqb1}u = \hat B(u) - 1/\hat
B(u)^3\end{equation} where $\hat B(u)  \equiv  B(u)/z$ and $\hat B(0)
= 1$. Unfortunately, integrating Eq. \ref{asymptotic_f} numerically
using Eq. \ref{eqb1}  to determine $f(u)$, we find that the resulting
``cross-integral" $\int_0^\infty ~B(u)~f(u) ~du \simeq 1.15$ is
somewhat larger than $1$.   This implies that a smaller value of the
rescaling factor $a$ must be used.  By carrying out the appropriate
numerical integrals we find that $a \simeq 0.70$  satisfies the
``cross-integral" normalization  condition  (\ref{crossintegral}).
The corresponding scaled CND and ISD with $b = 1$ are shown in Fig. 
1. As can be
seen, the calculated asymptotic ISD  is shifted to the right from the
simulation result and is cut off to zero at $u_c \simeq 1.85$ where
$C(u)$ crosses $z u$.
  Thus we find that, contrary to the claims in
Ref. \onlinecite{Coment03}, our method leads to a {\it non-divergent}
ISD in the asymptotic limit of infinite $D/F$.

\begin{figure}[!htb]
\includegraphics [width=8.5cm] {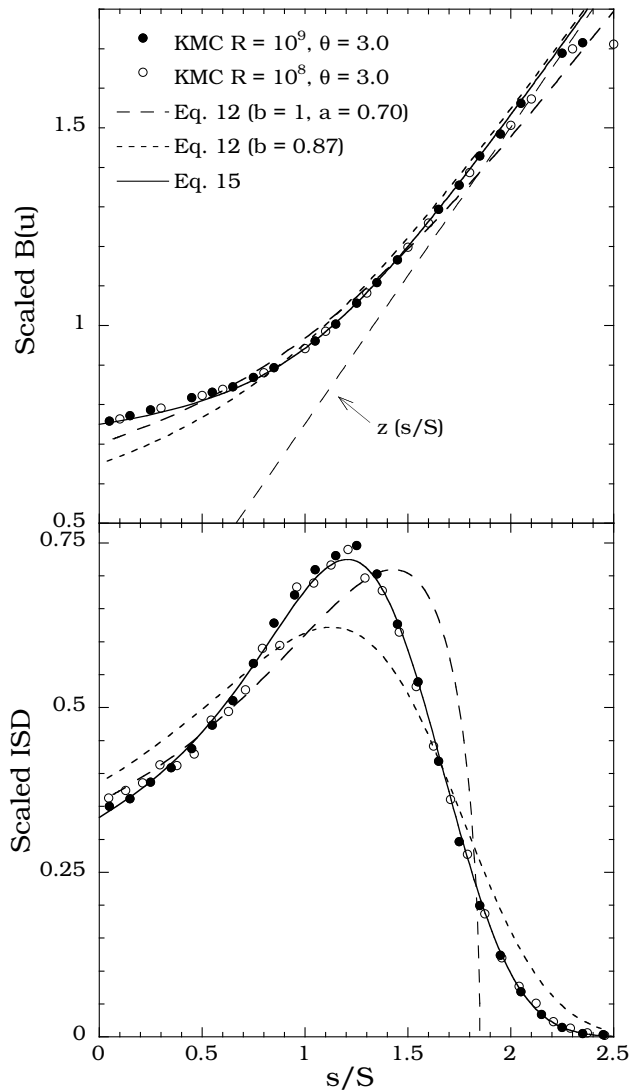}
\caption{\label{fig1}
Calculated asymptotic scaled (a)
gap-distribution $B(u)$ 
and (b) island-size distribution $f(u)$,
along with KMC results (symbols).
}
\end{figure}

It is also interesting to  consider  the general case $b \neq 1$,
i.e. the capture-area of a freshly-nucleated dimer is proportional to
the average area per island with some unknown proportionality
constant $b$.  In this case, we assume that the value of the
rescaling constant $a$ is  fixed to the value $a = z$ by the
requirement that the asymptotic behavior of $B(u)$ is given by $B(u)
\simeq z u$ for large $u$. We then search for a value of $b$ ($b
\simeq 0.87$) such that the sum-rule (\ref{crossintegral}) is
satisfied.  As can be seen in Fig. 1, the resulting scaled
gap-distribution  $B(u)$ is now very close to the simulation results
for all $u$ except for small $u$ ($u < 0.7$) where it is now
significantly lower than for the case $b = 1$.  Accordingly, the peak 
of the corresponding
island-size distribution (Fig. 1(b)) is somewhat lower than the peak
of the  simulated distribution, and the distribution itself is
somewhat wider.

We now present an
analytical form for the
capture-number distribution $C(u) = B(u)$
which provides excellent agreement with
simulations and thus strongly supports our
conjecture that $B(0) = z$ and that $B(u) \rightarrow zu$ for large $u$.
Since  Eq.~\ref{firstfix}
with $b = 1$ and $a  = z$ leads to an expression  for $B(u)$
(\ref{eqb1})
which satisfies $\hat B(0) = 1$ and $B(u) \rightarrow z u$ for large $u$,
but  a cross-integral (\ref{crossintegral}) which is just slightly
larger  than $1$,  we propose that the correct analytic expression
for $B(u)$ has the same form as (\ref{eqb1}) but with additional
terms corresponding to  higher order powers of   $[\hat B(u)]^{-1}$.
As an example, we chose the following form, \begin{equation}u = \hat
B(u) - (b_2 /\hat B(u)^{3})~ \exp[b_1/\hat
B(u)^n],\label{Cu_exp}\end{equation} where $b_2 = e^{-b_1}$ in order
to satisfy $\hat B(0) = 1$, and $n$ is a free parameter for which we
will choose $n = 5$. We then use Eq. \ref{asymptotic_f} to calculate
$f(u)$ and search for a value of $b_1$ ($b_1  \simeq 0.993$) such that
the sum-rule (\ref{crossintegral}) is satisfied. Our results for the
corresponding  $B(u)$ and $f(u)$ obtained in this manner are shown by
the solid  curves in Fig.~1. As can be seen, the agreement between
the calculated scaled gap-distribution $B(u)$  and simulations is now
excellent.  Accordingly, the agreement between the corresponding
calculated ISD  and simulations is also very good.
This result also confirms that $C(u) = B(u)$ in the asymptotic limit.

We have also tried several other values of $n$
($n = 3, 4, 6$) and obtained similarly  good agreement with
simulations.\cite{agreement}
Finite expansions in powers of $[\hat B(u)]^{-1}$ satisfying
$\hat B(0) = 1$ also lead to  reasonably good  agreement.  In particular, the
expression
\begin{equation}
u = \hat B(u) - \gamma \hat B(u)^{-3} - \frac{1 - \gamma}{2}
[\hat B(u)^{-9} + \hat B(u)^{-12}],
\end{equation}
(with $\gamma \simeq 0.39$ to satisfy (\ref{crossintegral}))  leads  to
results which are essentially identical to those obtained using
Eq.~\ref{Cu_exp}. Thus, it would appear that almost any form  
similar to
(\ref{eqb1}), but with   higher order corrections,
which satisfies the condition
$\hat B(0) = 1$ as well as the
sum-rule (\ref{crossintegral}), leads to
reasonably accurate results for the asymptotic scaled ISD and CND.

In conclusion, we have derived analytical expressions for the
asymptotic scaled capture-number, gap-, and island-size distributions
for the case of  one-dimensional irreversible growth of point
islands.
In particular, we have shown that our method
leads to an asymptotic scaled CND which is close to that obtained
in simulations and
as a result
   to a {\it non-divergent} asymptotic ISD.
Our analytical results  also  indicate that in the asymptotic
limit the scaled gap-distribution $B(u)$ is identical to the scaled
capture-number distribution $C(u)$.
Finally, by slightly modifying our  analytical expression
(\ref{eqb1}) for the scaled CND and solving self-consistently, we have
   obtained analytic expressions for the scaled  CND and
corresponding ISD which are in excellent agreement with simulations.

We note that our analytical results
also demonstrated  that
the ``self-averaging property" of the capture zones  (i.e. the 
average capture number  $\sigma_s$
is exactly equal to the
local capture number evaluated at the average capture zone of islands 
of size $s$)
is   exact  in the asymptotic limit for the case of
irreversible growth of point-islands in one-dimension.  If this
``self-averaging property" holds generally (as  suggested by the good
agreement obtained in our numerical results for finite $D/F$ in both
one- and two-dimensions \cite{APFsurf, JMF_recent}) then it may be
possible in general to obtain  accurate asymptotic  ISDs without having to
know
the exact distribution of areas or the
redistribution of areas due to nucleation events.
This
   represents a
significant simplification of the calculation
of ISDs since the determination of the full distribution of capture
zones is a very difficult problem for which no exact solution seems
to be possible even in the most simple cases.\cite{BE_recent,recent}

Finally, we consider the extension of these results
to other cases of interest such as growth in
two-dimensions. The present analytical work was essentially
dependent on the asymptotically valid relation $B(u) = C(u)$.
It is not clear  that such an explicit relation may be derived for 
the case of two-dimensional growth,
and additional mathematical
difficulties occur due to complicated expressions for the local
capture numbers. As a result,  it is not clear if  a straightforward
extension of the present work to this case  is   possible, and further
work is required to answer this question.

J. G. Amar would like to acknowledge support  from the Petroleum Research Fund
and from the NSF (DMR-0219328).


\begin{thebibliography}{}

\bibitem{scaling}
M.C. Bartelt and J.W.Evans,
Phys. Rev. B {\bf 46}, 12675 (1992);
C. Ratsch, A. Zangwill, P. Smilauer, and D.D. Vvedensky,
Phys. Rev. Lett. {\bf 72}, 3194 (1994);
J.G. Amar, F. Family, and P.M. Lam,
Phys. Rev. B {\bf 50}, 8781 (1994);
M.C. Bartelt and J.W. Evans,
J. Vac. Sci. Tech.  A {\bf 12}, 1800 (1994);
J.G. Amar and F. Family,
Phys. Rev. Lett. {\bf 74}, 2066 (1995);
P.A. Mulheran and J.A. Blackman,
Philos. Mag. Lett. {\bf 72}, 55 (1995).

     \bibitem{BErapid}
M.C. Bartelt and J.W. Evans,
Phys. Rev. B {\bf 54}, R17359 (1996).  

\bibitem {re_sc_1d}
J.A. Blackman and P.A. Mulheran,
Phys. Rev. B {\bf 54}, 11681 (1996);
P.A. Mulheran and J.A. Blackman,
Surf. Sci. {\bf 376}, 403 (1997).

\bibitem{recent}
P.A. Mulheran and D.A. Robbie,
Europhys. Lett. {\bf 49}, 617 (2000);
F.G. Gibou, C. Ratsch, M.F. Guyre, S. Chen, R.E. Caflisch,
Phys. Rev. B {\bf 63}, 115401 (2001);

\bibitem{Vvedensky}D.D. Vvedensky,Phys. Rev. B {\bf 62}, 15435 (2001).

\bibitem{JMF_recent}
J.G. Amar, M.N. Popescu, and F. Family,
Phys. Rev. Lett. {\bf 86}, 3092 (2001);
M.N. Popescu, J.G. Amar, and F. Family,
Phys. Rev. B {\bf 64}, 205404 (2001).

\bibitem{APFsurf}
J.G. Amar, M.N. Popescu, and F. Family,
Surf. Sci. {\bf 491}, 239 (2001).

\bibitem{BE_recent}
J.W. Evans and M.C. Bartelt,
Phys. Rev. B {\bf 63}, 235408 (2001);
\textit{ibid.} {\bf 66}, 235410 (2002).

\bibitem {Smo}
M. von Smoluchowski,
Z. Phys. Chem. {\bf 17}, 557 (1916);
\textit{ibid.} {\bf 92}, 129 (1917).

\bibitem{Venables}
J.A. Venables,
Philos. Mag. {\bf 27}, 697 (1973); J.A. Venables, G.D. Spiller, and 
M. Hanbucken,
Rep. Prog. Phys. {\bf 47}, 399 (1984).


\bibitem{Coment03}
D.D. Vvedensky, C. Ratsch, F. Gibou, and R. Vardavas,
Phys. Rev. Lett. {\bf 90}, 189601 (2003);
J.G. Amar, M.N. Popescu, and F. Family,
Phys. Rev. Lett. {\bf 90}, 189602 (2003).


\bibitem{note1}
Our model of  irreversible growth of point-islands in one-dimension
may be defined as follows. Atoms are deposited randomly on an
initially
empty line of sites with a (per site) deposition rate $F$ and
diffuse with a diffusion constant $D$. When a monomer moves onto
a site occupied by another monomer, a dimer island is nucleated
at that site. Similarly, a monomer moving onto a site occupied
by an island is absorbed and the island size $s$ increases
by one. All  islands of size $s \geq 2$ are assumed to be stable
and immobile.


\bibitem{largey} In the asymptotic limit a vanishingly small fraction
of islands will have nucleated at a coverage $\theta_y$ which does
not satisfy $\theta_y >> \theta_x$ and therefore this assumption
holds for essentially all islands. \bibitem {agreement} For example,
Eq. \ref{Cu_exp} with $n = 3$ leads to a distribution with a slightly
higher peak which is just slightly shifted to the right and with a
tail which is just slightly lower than in simulations.

\end{thebibliography}
\end{document}